\documentclass[aps,prl,twocolumn,amsmath,amssymb,superscriptaddress,floatfix]{revtex4}
\usepackage{graphicx}
\usepackage{bm}
\usepackage[usenames]{color}
\usepackage{colordvi}
\bibstyle{apsrev.bib}

\newcommand{\be}{\begin{equation}}
\newcommand{\ee}{\end{equation}}
\newcommand{\beqn}{\begin{eqnarray}}
\newcommand{\eeqn}{\end{eqnarray}}

\begin{document}

\title{Quantum relaxation after a quench in systems with boundaries}
\author{Ferenc Igl\'oi}
\email{igloi@szfki.hu}
\affiliation{Research Institute for Solid State Physics and Optics,
H-1525 Budapest, P.O.Box 49, Hungary}
 \affiliation{Institute of Theoretical Physics,
Szeged University, H-6720 Szeged, Hungary}
\author{Heiko Rieger}
\email{h.rieger@mx.uni-saarland.de}
\affiliation{Theoretische Physik, Universit\"at des Saarlandes, 66041 Saarbr\"ucken, Germany}
\date{\today}

\begin{abstract}
We study the time-dependence of the magnetization profile, $m_l(t)$,
of a large finite open quantum Ising chain after a quench.  We observe
a cyclic variation, in which starting with an exponentially decreasing
period the local magnetization arrives to a quasi-stationary regime,
which is followed by an exponentially fast reconstruction period.  The
non-thermal behavior observed at near-surface sites turns over to
thermal behavior for bulk sites. Besides the standard time- and
length-scales a non-standard time-scale is identified in the
reconstruction period.
\end{abstract}


\maketitle

Recent experimental progress in controlling ultracold atomic gases in
optical lattices has opened new perspectives in the physics of quantum
systems. In these measurements the coupling in an interacting system
can be tuned very rapidly, commonly denoted as ``quench'', for
instance by using the phenomenon of Feshbach resonance and the
couplings to dissipative degrees of freedom (such as phonons and
electrons) are very weak. As a consequence one can study coherent time
evolution of isolated quantum systems. Among the fascinating new
experiments we mention the collapse and revival of Bose-Einstein
condensates\cite{BE_exp}, quenches in a spinor
condensate\cite{spinor}, realization of one-dimensional Bose
systems\cite{1d} and measurements of their non-equilibrium
relaxation\cite{1d_dyn}.

Concerning the theoretical side of quantum quenches here the first
investigations had been performed on quantum XY and quantum Ising spin
chains\cite{barouch_mccoy,igloi_rieger,sengupta} before the
experimental work has been started.  The new experimental results in
this field have triggered intensive and systematic theoretical
researches, which are performed on different systems, such as 1D Bose
gases\cite{rigol}, Luttinger liquids\cite{cazalilla} and
others\cite{manmana}. Besides studies on specific models there are
also field-theoretical investigations, in which relation with boundary
critical phenomena and conformal field-theory are
utilized\cite{calabrese_cardy,sotiriadis_cardy}.

One fundamental question of quantum quenches concerns the nature of
the stationary state of this non-equilibrium quantum relaxation
including the issue of thermalization and potential descriptions 
by Gibbs ensembles.
For non-integrable systems exact thermalization of stationary states
was conjectured\cite{gritsev}, however the numerical results on
specific systems are controversial\cite{gritsev,roux,kollath}. 
On the other hand integrable systems are sensitive to the initial states and
their stationary states are thermal-like being in a form of a
generalized Gibbs ensemble\cite{rigol}.

Thermalization includes generically (i.e. away from critical points)
an exponential decay of correlation functions in the stationary
state on length and time scales that can be
related to the correlation length and time of an equilibrium system at
an effective temperature depending on the parameters of the quench
\cite{sengupta,rossini,barmettler}. Some quantum systems do 
not thermalize completely and display a different behavior for
correlation functions of local and for non-local operators, such that
the former do not exhibit effective thermal behavior\cite{rossini}. An
interesting issue not being addressed so far is the characterization
of the non-stationary, that means not time-translation invariant,
quantum relaxation following a quench: Preparing the quantum system in
a non-eigenstate of its Hamiltonian, how is thermalization achieved
during the time-evolution? How do correlations develop in time towards
the stationary (i.e.\ time translation invariant) state, is there
a time dependent correlation length, etc.?

Another important issue concerns quantum relaxation and potential
thermalization in the presence of boundaries.  Theoretical studies of
non-equilibrium quantum relaxation have focused on bulk sites up to now,
but all real systems have a finite extent and they are bounded by
surfaces and the physical properties in the surface region are considerably
different from those in the bulk\cite{binder83}. Obviously an interesting
question is whether the time and length scales characterizing the 
stationary relaxation in the bulk is altered in the vicinity of
the boundary, and how thermalization is achieved there.

In this paper we will address these two issues: The non-stationary
quantum relaxation after a quench and the effect of boundaries.
For this we focus on a computationally tractable model for a quantum 
spin chain and study the relaxation of profiles of observables in
the early time steps as well as their behavior in the long-time
limit. We also address the behavior in large, but finite
systems and study the consequences of the recurrence
theorem.

The system we consider in this paper is the quantum Ising chain
defined by the Hamiltonian:
\be
{\cal H}=-\sum_{l=1}^{L-1} \sigma_l^x \sigma_{l+1}^x -h \sum_{l=1}^{L} \sigma_l^z\;,
\label{hamilton}
\ee
in terms of the Pauli-matrices $\sigma_l^{x,z}$ at site $l$.  In the
non-equilibrium process the strength of the transverse field is
suddenly changed from $h_0$ ($t<0$) to $h$ ($t \ge 0$).  The
Hamiltonian in Eq.(\ref{hamilton}) can be expressed in terms of free
fermions\cite{pfeuty}, which is used in studies of its non-equilibrium
properties\cite{igloi_rieger,rossini}.  
The bulk transverse magnetization,
$\sigma_l^z$, which is a local operator, has non-thermal behavior\cite{barouch_mccoy,calabrese_cardy,barmettler},
whereas the bulk (longitudinal) magnetization, $\sigma_l^x$, which is
a non-local operator, has effective thermal behavior\cite{barmettler}.  Here we
concentrate on the latter quantity and study the time-dependence of
its profile,
$m_l(t)=\lim_{b\to0_+}\, _b\langle \varPsi^{(0)}_0 | \sigma_l^x(t) |
\varPsi^{(0)}_1 \rangle_b$, where $|\varPsi^{(0)}_0\rangle_b$ is the
ground state of the initial Hamiltonian (\ref{hamilton}) in the
presence of an external longitudinal field $b$.
According to
\cite{yang} this can be written as the off-diagonal matrix-element
of the Hamiltonian (\ref{hamilton}):
\be
m_l(t)=\langle \varPsi^{(0)}_0 | \sigma_l^x(t) | \varPsi^{(0)}_1 \rangle\;.
\label{offd_magn}
\ee
Here $|\varPsi^{(0)}_1\rangle$ is the first excited state (which is
the ground state of the sector with odd number of fermions) of the
initial Hamiltonian ($t<0$). In the ordered phase, $h_0<h_c=1$ where
$m_l(t<0)>0$, $|\varPsi^{(0)}_1\rangle$ is asymptotically degenerate
with the ground state, $|\varPsi^{(0)}_0\rangle$. For $h_0 \ge h_c$
the magnetization vanishes as $m_l(t<0) \sim L^{-x}$ with the system
size for $t<0$.  The decay exponent, $x$, is different at the critical
point, $h=h_c$, and in the paramagnetic phase, $h>h_c$, as well in the
bulk ($l/L=O(1)$) and at the boundary ($l/L \to 0$), see
Table~\ref{table:1}.

\begin{table}
\caption{Decay exponent of the off-diagonal (longitudinal) magnetization in the initial (equilibrium) period.\label{table:1}}
 \begin{tabular}{|c|c|c|}  \hline
   & $h_0=h_c$ & $h_0>h_c$ \\ \hline
bulk   &  $1/8$  & $1/2$ \\
boundary   & $1/2$  & $3/2$ \\ \hline
  \end{tabular}
  \end{table}

To calculate the magnetization profile in Eq.(\ref{offd_magn}) we have
used standard free-fermionic techniques\cite{lieb,pfeuty}.
For the surface site, $l=1$, most of the calculations are 
analytical, whereas for $l>1$ numerical calculations
have been made for large finite systems up to $L=384$.

\begin{figure}[h!]
\begin{center}
\includegraphics[width=1.6in,angle=0]{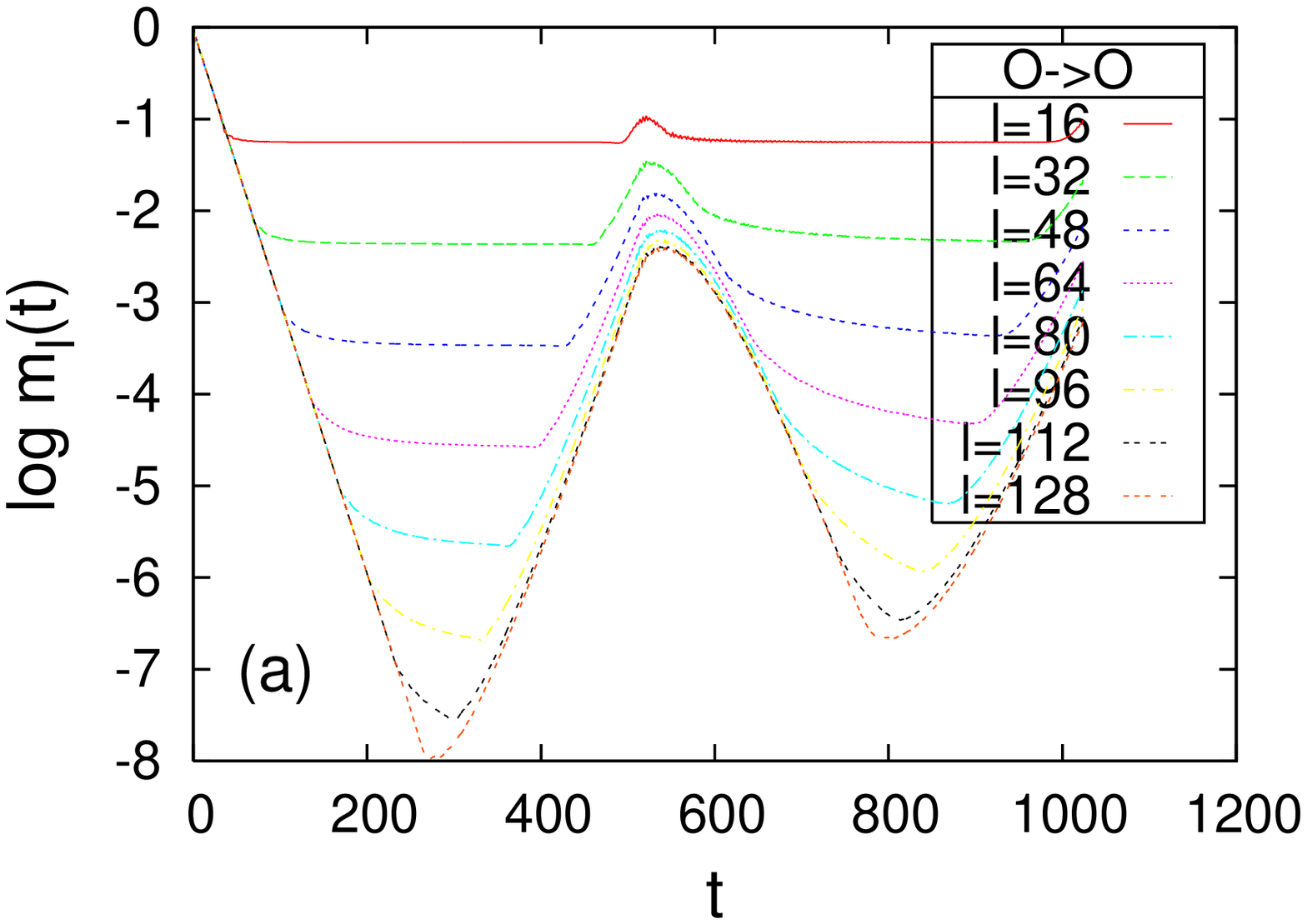}
\includegraphics[width=1.6in,angle=0]{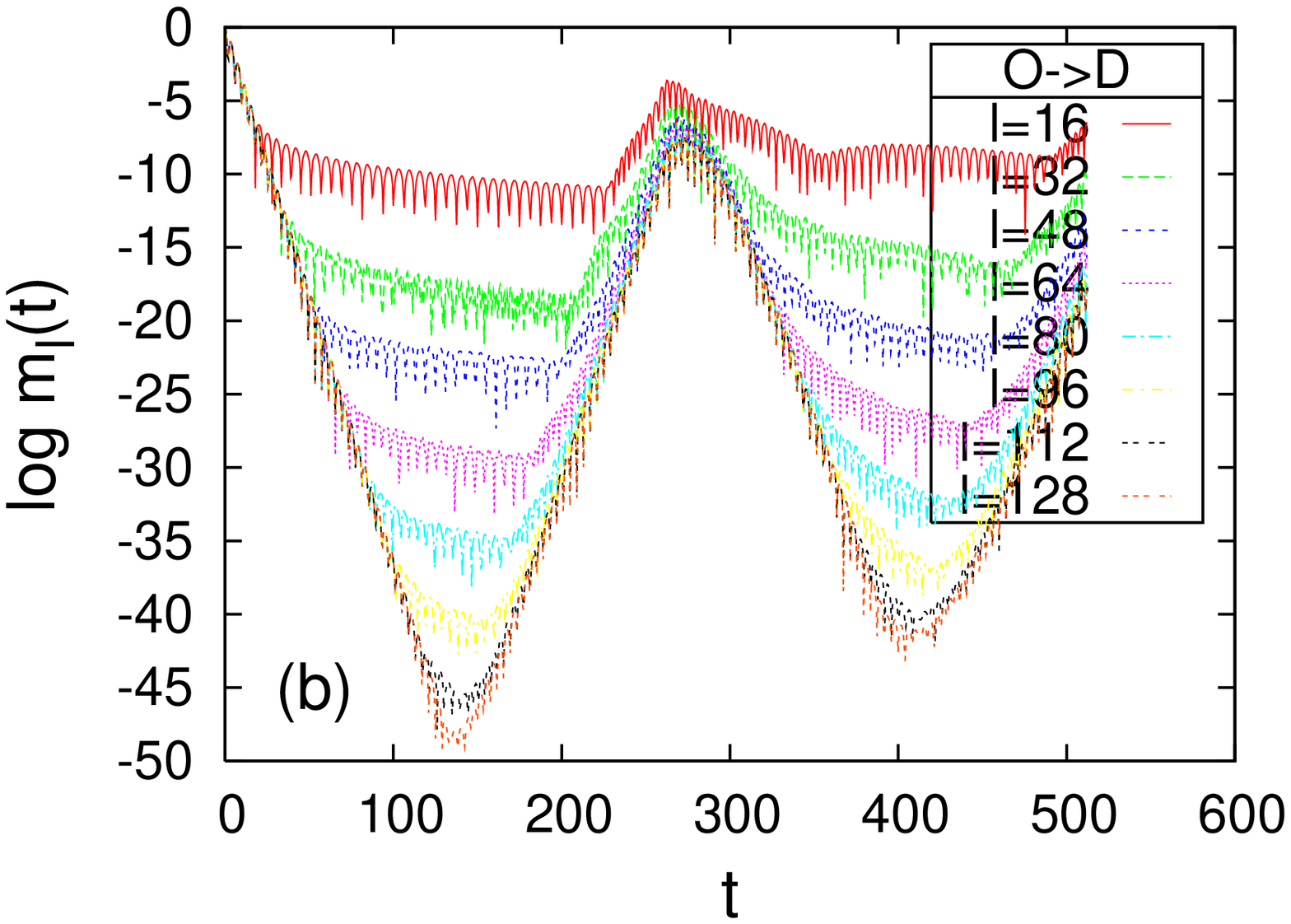}
\includegraphics[width=1.6in,angle=0]{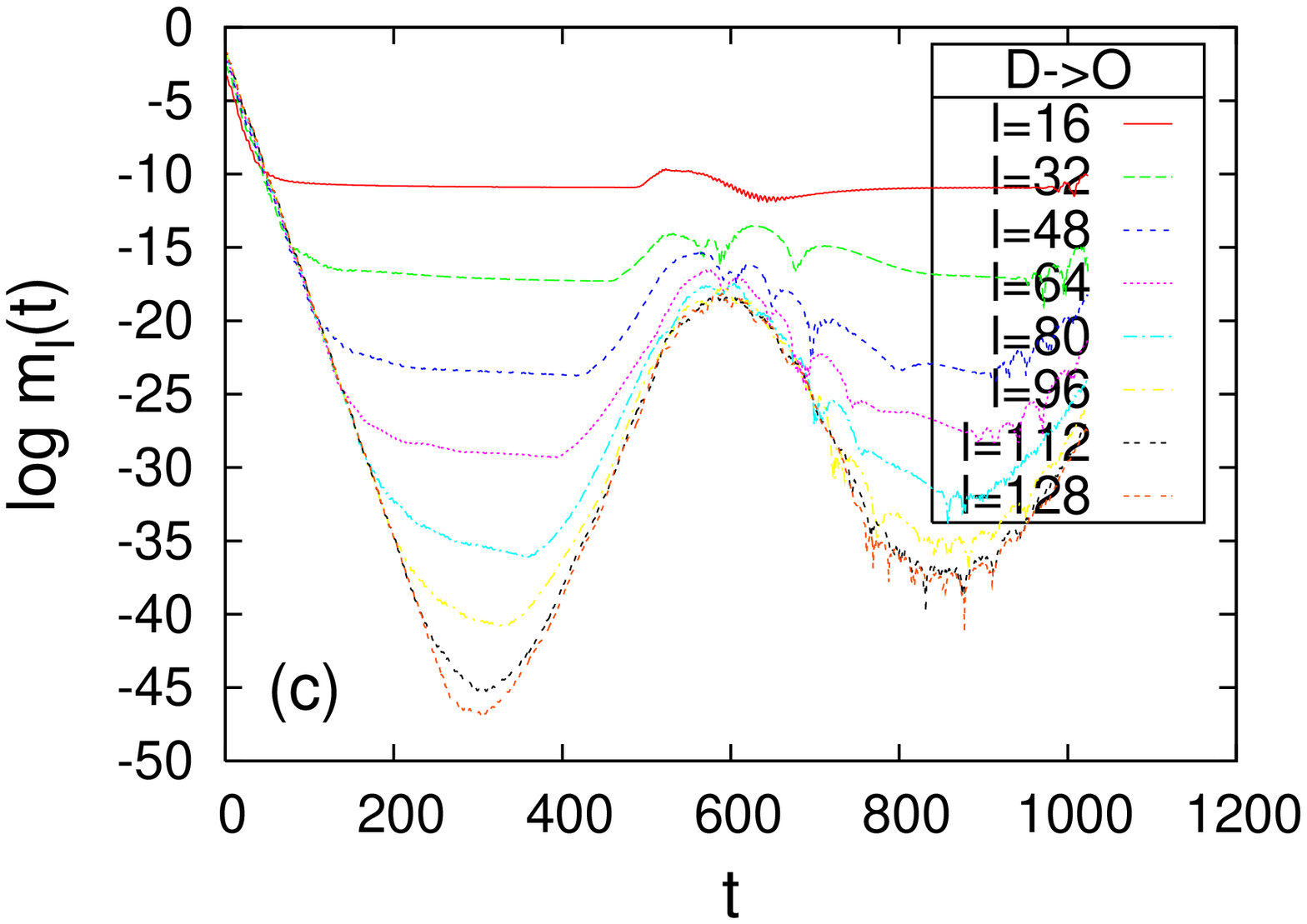}
\includegraphics[width=1.6in,angle=0]{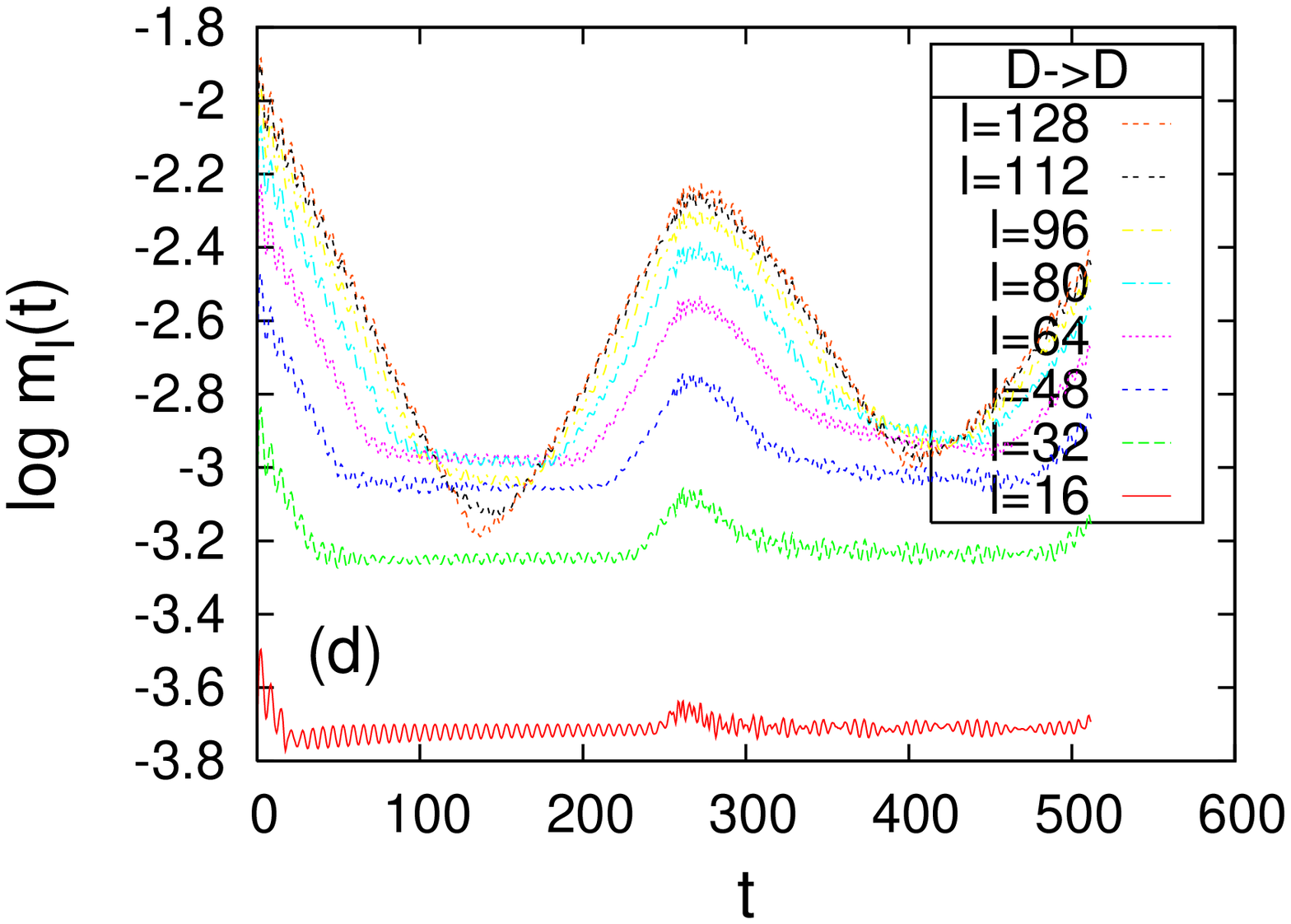}
\end{center}
\vskip -.5cm
\caption{
\label{fig_1} (Color online)
Relaxation of the local magnetization, $\log m_l(t)$, at different positions in a $L=256$ chain after a quench with
parameters: a) $h_0=0.0$ and $h=0.5$ (\textbf{O} $\rightarrow$ \textbf{O}) b) $h_0=0.5$ and $h=1.5$ (\textbf{O} $\rightarrow$ \textbf{D})
c)  $h_0=1.5$ and $h=0.5$ (\textbf{D} $\rightarrow$ \textbf{O})d)  $h_0=1.5$ and $h=2.0$ (\textbf{D} $\rightarrow$ \textbf{D}).
}
\end{figure}

We have performed quenches for various pairs of transverse fields,
$h_0$ and $h$ and calculated the time-dependence of the local
magnetization at different sites,
$l \le L/2$.  The results depend primarily on whether the system
before and after the quench is in the ordered (\textbf{O}) or
disordered (\textbf{D}) phase, see Fig.\ref{fig_1} for different
combinations of \textbf{O} and \textbf{D}. One can identify different
time regimes that can be interpreted in terms of quasi-particles,
which are emitted at $t=0$,
travel with a constant speed, $v=v(h,h_0)$, and are reflected at the
boundaries.

As argued in Ref.\cite{calabrese_cardy} only those quasi-particles are
quantum entangled that originate from nearby regions in space, others
are incoherent. When the latter arrive at a reference point $l$ they
cause relaxation of local observables (such as magnetization). Here we
extend this picture by noting that in a system with boundaries the
same quasi-particle can reach the point $l$ twice (or more) at
different times after reflections. This induces quantum correlations
in time signalized by the reconstruction of the value of the local
observable. In the following we analyze the different regimes of the
relaxation.

In the \underline{free relaxation regime: $t < t_l=l/v$}
only incoherent quasi-particles pass the reference point 
resulting in an exponential decay of the magnetization 
(cf. Fig.\ \ref{fig_1}):
\be
m_l(t) \equiv m(t) \approx A(t) \exp(-t/\tau),\quad t<t_l \;,
\label{rel_time}
\ee
with an oscillating prefactor, $A(t)$.  In the regime $h>h_c$ and
$h_0<h_c$ we have $A(t) \sim \cos(at+b)$, thus $m(t)$ changes sign. On
the other hand in the other parts of the phase diagram $m(t)$ is
always positive, i.e. $A(t) \sim [\cos(at+b)+c]$, with $c>1$.  The
characteristic time-scale, $\tau=\tau(h,h_0)$, is the relaxation or
phase coherence time, which is extracted from the numerical data. The
exponential form of the decay in Eq.(\ref{rel_time}) indicates
thermalization, at least for bulk sites, which is in agreement with
the similar decay of the autocorrelation function.

In the \underline{quasi-stationary regime: $t_l<t<T-t_l$ }, $T=L/v$, 
two types of quasi-particles reach the reference point $l$: type \textit{1}
passed $l$ only once at a time $t'<t$ and type \textit{2} passed it twice
at two times $t'<t''<t$ with a reflection at the nearby boundary
between $t'$ and $t''$. These two types interfere, resulting in a
comparatively slow relaxation (cf. Fig.\ \ref{fig_1}). Deep inside
the ordered phase the quasi-particles can be identified with kinks
moving with a speed $\pm v$ \cite{sachdev_young} and in the regime
$t_l\ll t\ll T$ half of the quasi-particles reaching the site $l$ are of
type \textit{1} (flipping the spin at $l$ once) and half of them type \textit{2}
(flipping it twice), leading to a quasi-stationary relaxation.

\begin{figure}[t]
\begin{center}
\includegraphics[width=1.6in,angle=0]{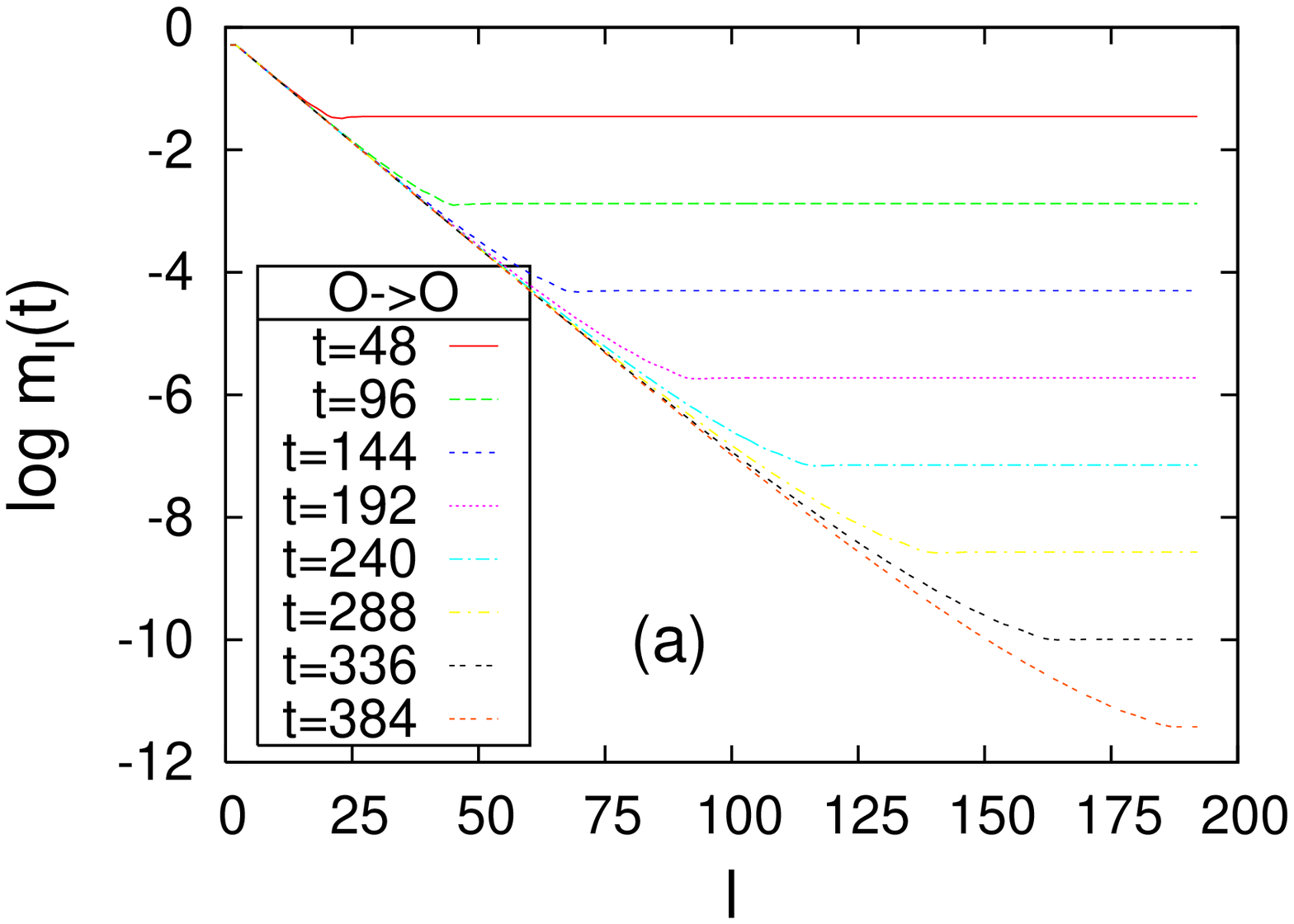}
\includegraphics[width=1.6in,angle=0]{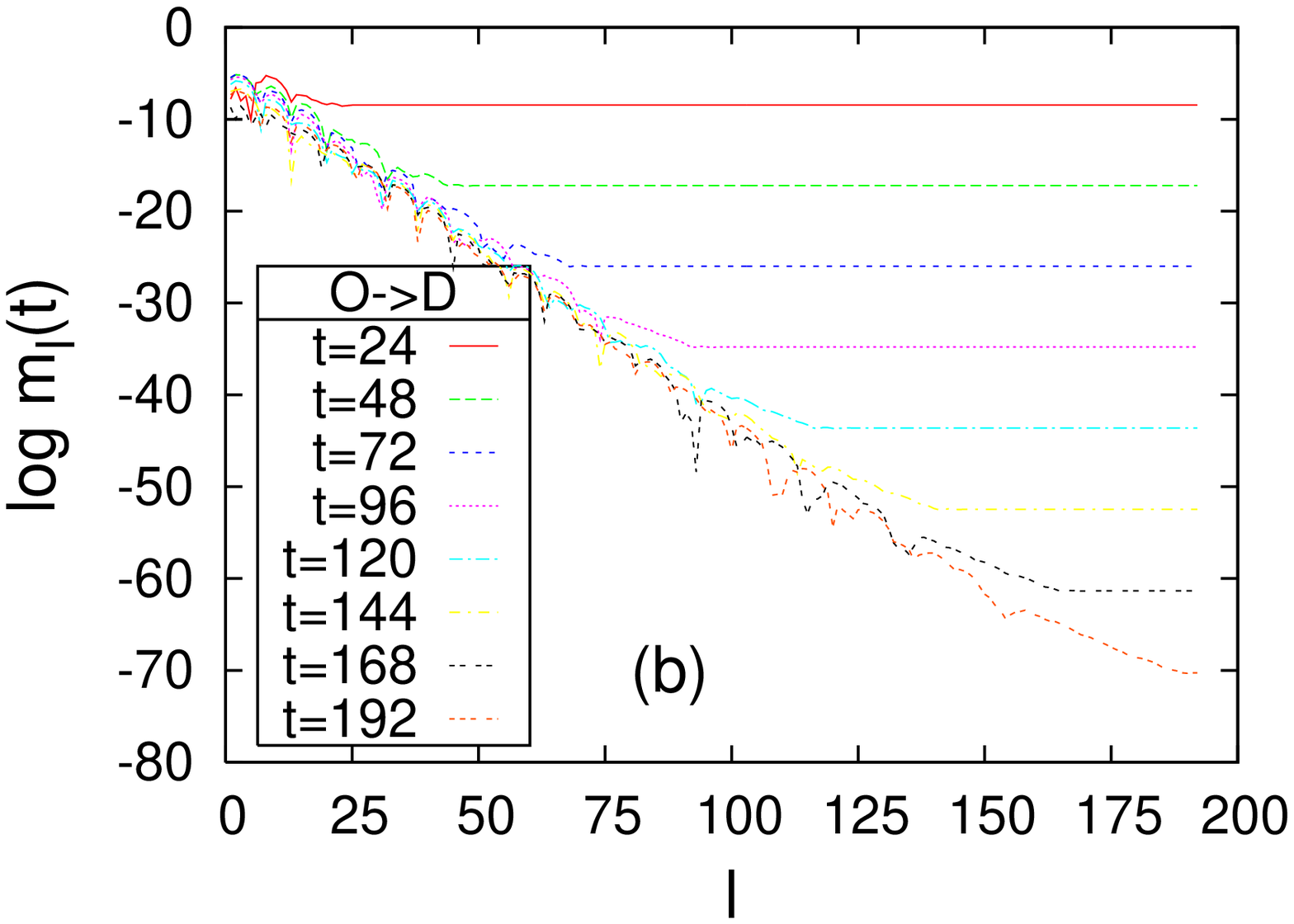}
\includegraphics[width=1.6in,angle=0]{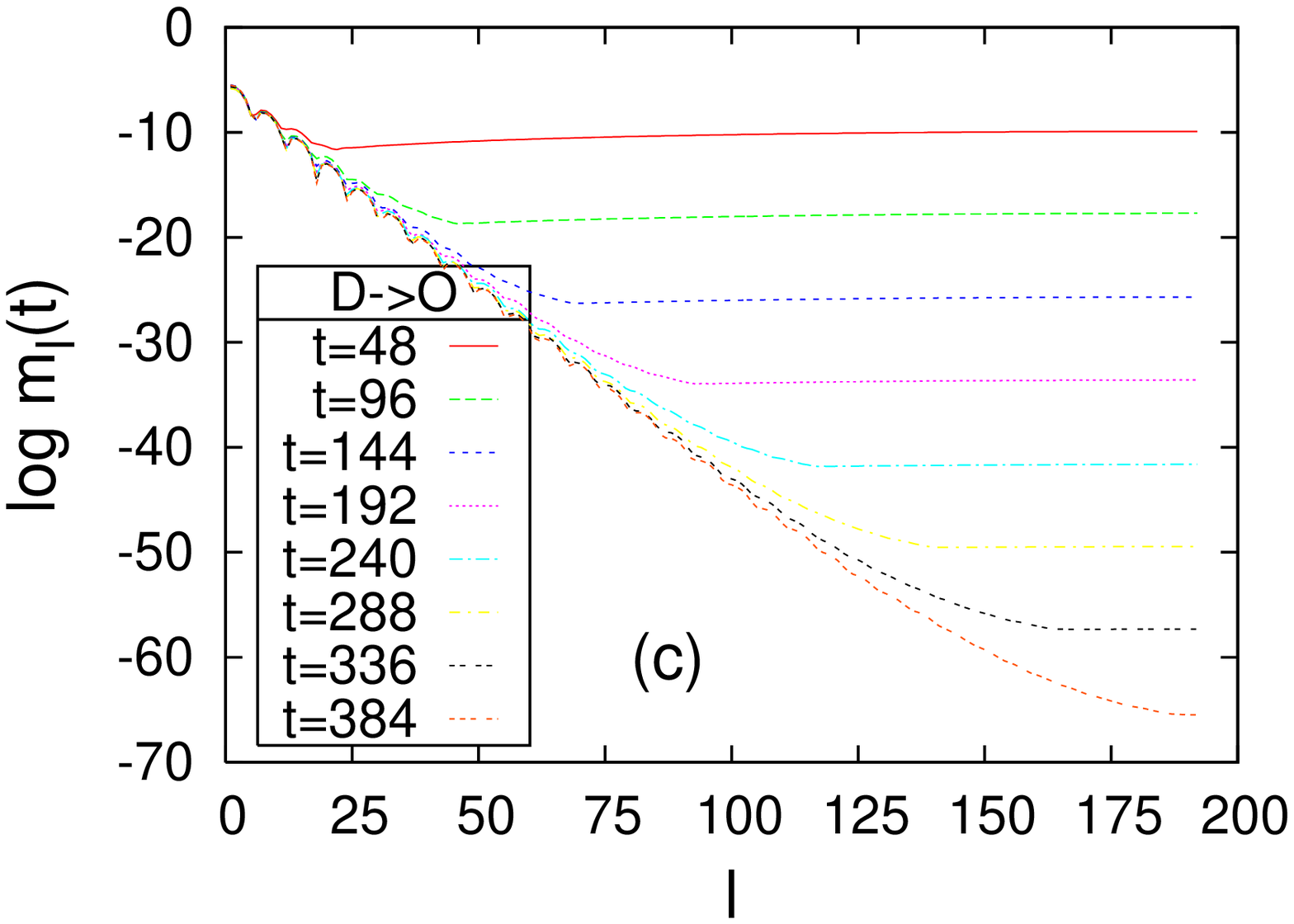}
\includegraphics[width=1.6in,angle=0]{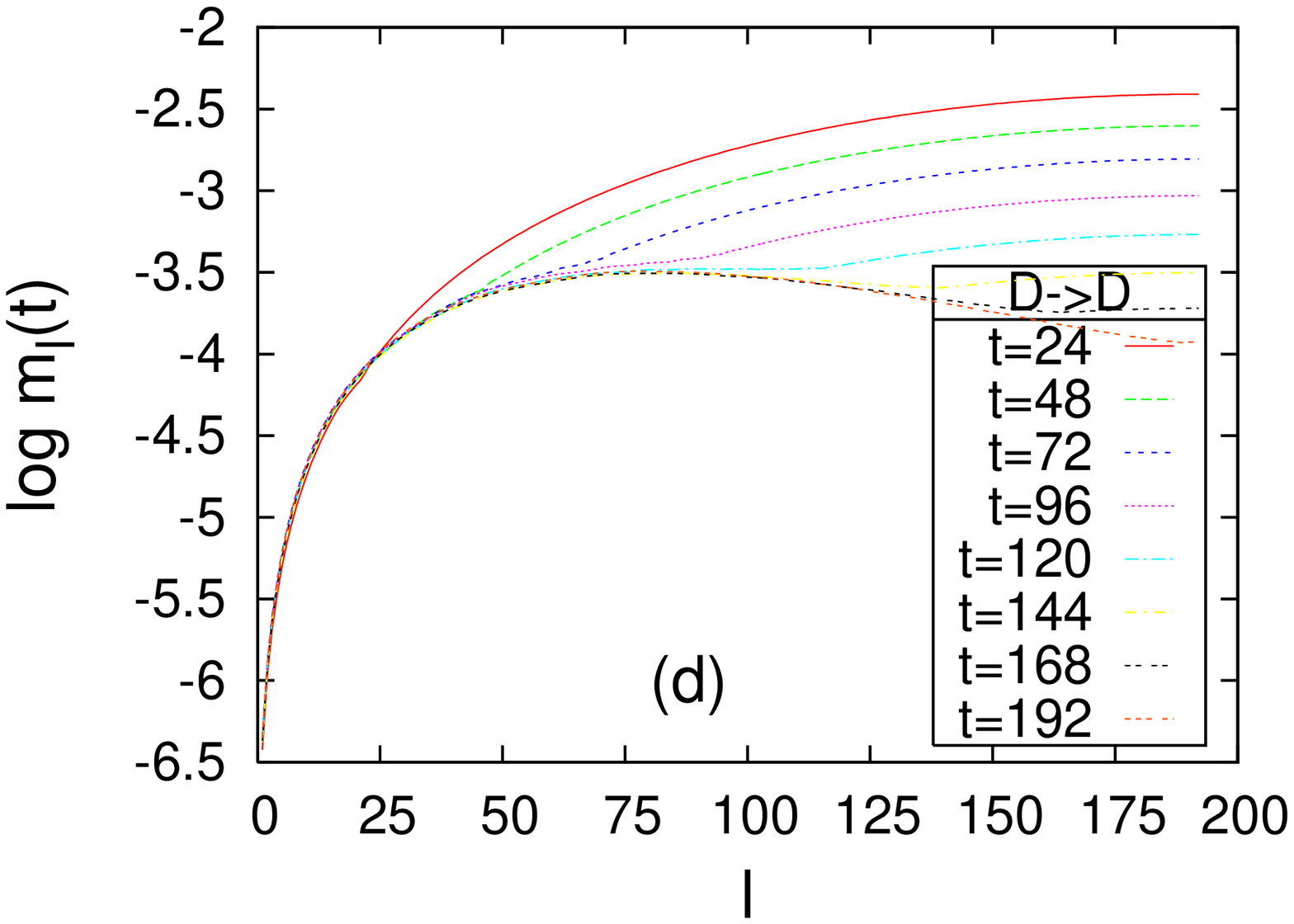}
\end{center}
\vskip -.5cm
\caption{
\label{fig_2} (Color online)
Non-equilibrium magnetization profiles, $\log m_l(t)$, at different times after a quench with parameters
given in Fig.\ref{fig_1} for $L=384$. From the asymptotic values of the slopes one can measure the
correlation length.
}
\end{figure}

The magnetization profiles for fixed times $t<T/2$ are shown in
Fig.\ref{fig_2} for the same quenches as in
Fig.\ref{fig_1}. For sufficiently large $l$ the quasi-stationary
magnetization has an exponential dependence, such that comparing its
value at two sites, $l_1$ and $l_2$, we have
\be
m_{l_1}(t_1)/m_{l_2}(t_2) \approx \exp\left[ -(l_1-l_2)/\xi\right] \;,
\label{xi}
\ee
with oscillating prefactors.

In the limits $L \to \infty$ and $t \to \infty$ one can define a
quasi-stationary limiting value which will be denoted by,
$\overline{m}_l$. For the surface site we have the exact result
\be
\overline{m}_1=\frac{(1-h^2)(1-h_0^2)^{1/2}}{1-h h_0},\quad h_0,h < 1\;,
\label{m1}
\ee
and zero otherwise. Note that the non-equilibrium surface magnetization
has different type of singularities for $h \to 1^{-}$ ($h_0<1$) and
for $h_0 \to 1^{-}$ ($h<1$).
We have analyzed the correction term,
$\Delta(t,L)=m_1(t)-\overline{m}_1$, and its asymptotic behavior is
summarized in Table~\ref{table:2} in the different domains of $h$ and
$h_0$. These corrections are in power-law
form, which signals that the relaxation of the surface magnetization
has non-thermal behavior.

\begin{table}
\caption{Correction to the quasi-stationary behavior for the surface magnetization in different domains of the quench.\label{table:2}}
 \begin{tabular}{|c|c|c|}  \hline
   & $h_0<h_c$ & $h_0>h_c$ \\ \hline
$h < h_0$   &  $t^{-1}\cos(at+b)$  & $L^{-3/2}[\cos(at+b)+c], \quad c>1$ \\
 $h > h_0$  & $t^{-3/2}\cos(at+b)$  & $t^{-1/2}[\cos(at+b)+cL^{-3/2}]$ \\ \hline
  \end{tabular}
  \end{table}

For $l>1$ we observe that $\overline{m}_{l}$ is monotonously
decreasing with $l$ and thus $\overline{m}_{l}>0$ for $h_0,h < 1$ and zero otherwise. The correction terms are identical with those given in 
Table~\ref{table:2} so that a finite
distance, $l$, the local magnetization has non-thermal behavior.

In the \underline{reconstruction regime: $T-t_l<t<T$} more and more
quasi-particles of type \textit{2} reach the reference point, which implies,
within a kink-picture, that incoherent spin flips in the past are
progressively reversed by quasi-particles returning to the site $l$
after reflection. For mono-disperse quasi-particles (velocity $v$) one
would expect a $T$-periodicity and thus $m_l(t)=m_l(T-t)$, i.e an
exponential increase in $t$ with a growth rate similar to the initial
decay rate. Indeed we find
\be
m_l(t) \equiv m(t) \approx B(t) \exp(t/\tau'),\quad T-t_l<t < T \;,
\label{rel_time1}
\ee
which is practically position independent and where the growth rate of
$\tau'(h,h_0)$ depends on the conditions of the quench, being
approximately proportional to $\tau(h,h_0)$: $\tau/\tau'=0.883\pm
0.002$. It turned out to be useful to measure the cross-over time,
$\tilde{t}=T/2$, which is defined as the crossing point of the two
asymptotic regimes: $\overline{A}\exp(-\tilde{t}/\tau)=\overline{B}
\exp(\tilde{t}/\tau')$, where $\overline{A}$ and $\overline{B}$ are
averaged prefactors. During the cross-over time the quasi-particles
travel a distance, $L/2$, thus their speed is given by:
$v(h,h_0)=L/2\tilde{t}$, which can be measured accurately.  We have
noticed, that for $h<1$ the speed is proportional to $h$: $v(h,h_0)=h
a(h,h_0)$, where $a(h,h_0)$ is practically independent of $h_0$ and
has just a very week dependence on $h$ close to $h=1$. The typical
values are in the range $a(h,h_0) \approx 0.86-0.88$. For $h \ge 1$
the speed is practically constant and has no $h$ dependence.

\underline{Approximate periodicity with $T$} starts for $t>T$, 
when quasi-particles start to be reflected second time and the spin-configuration
of the system becomes approximately equivalent to that at $t-T$.

\underline{The time- and length scale}, as defined in
Eq.(\ref{rel_time}) and Eq.(\ref{xi}), respectively, as well as the
characteristic quasi-particle speed $v(h,h_0)=\xi/\tau$, can be
extracted with high numerical accuracy from our data for the
magnetization profiles, typically with a precision of $3-4$ digits.
Complementary calculations of the autocorrelation function
$G_l(t)=\langle\varPsi_0^{(0)}|\sigma_l^x(t)\sigma_l^x(0)|\varPsi_0^{(0)}\rangle$,
and the equal-time correlation function,
$C_t(r)=\langle\varPsi_0^{(0)}|\sigma_{l+r}^x(t)\sigma_l^x(t)|\varPsi_0^{(0)}\rangle$
show that they yield the same correlation time and length, but with
less accuracy.  Based on our results for the profiles we have
conjectured possibly exact results about the relaxation time, as
discussed below.

\underline{The relaxation time $\tau(h,h_0)$} is divergent at two points:
\textit{i)} at the stationary point, $h=h_0$, where
$\tau(h,h_0) \sim (h-h_0)^{-2}$ and \textit{ii)} for small $h$, where
$\tau(h,h_0) \sim h^{-1}$, which can be derived
perturbatively. For $h_0=0$ the two singularities merge at
$h=0$: $\tau(h,h_0=0) \sim h^{-3}$.

\begin{figure}[h!]
\begin{center}
\includegraphics[width=1.6in,angle=0]{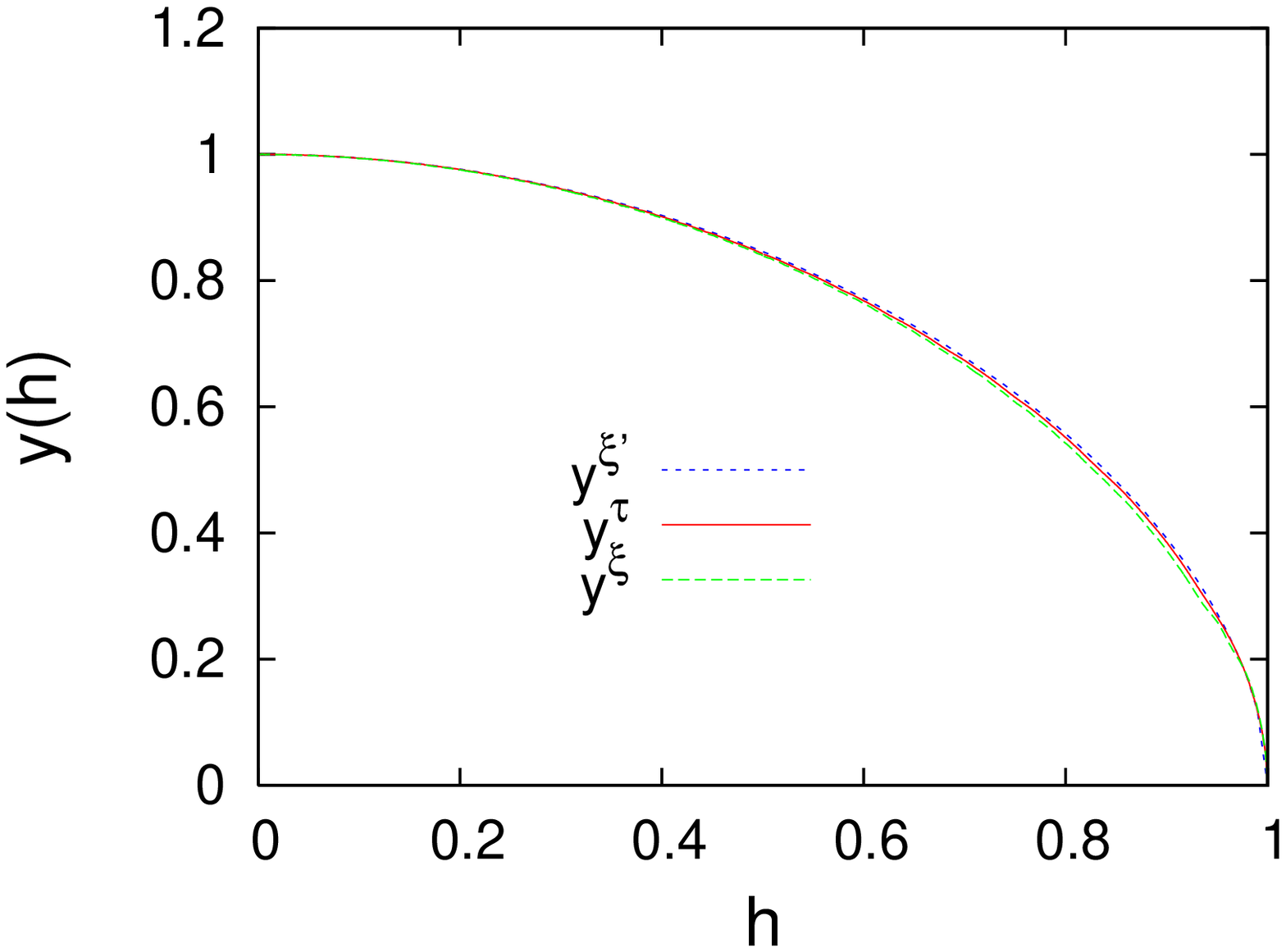}
\includegraphics[width=1.6in,angle=0]{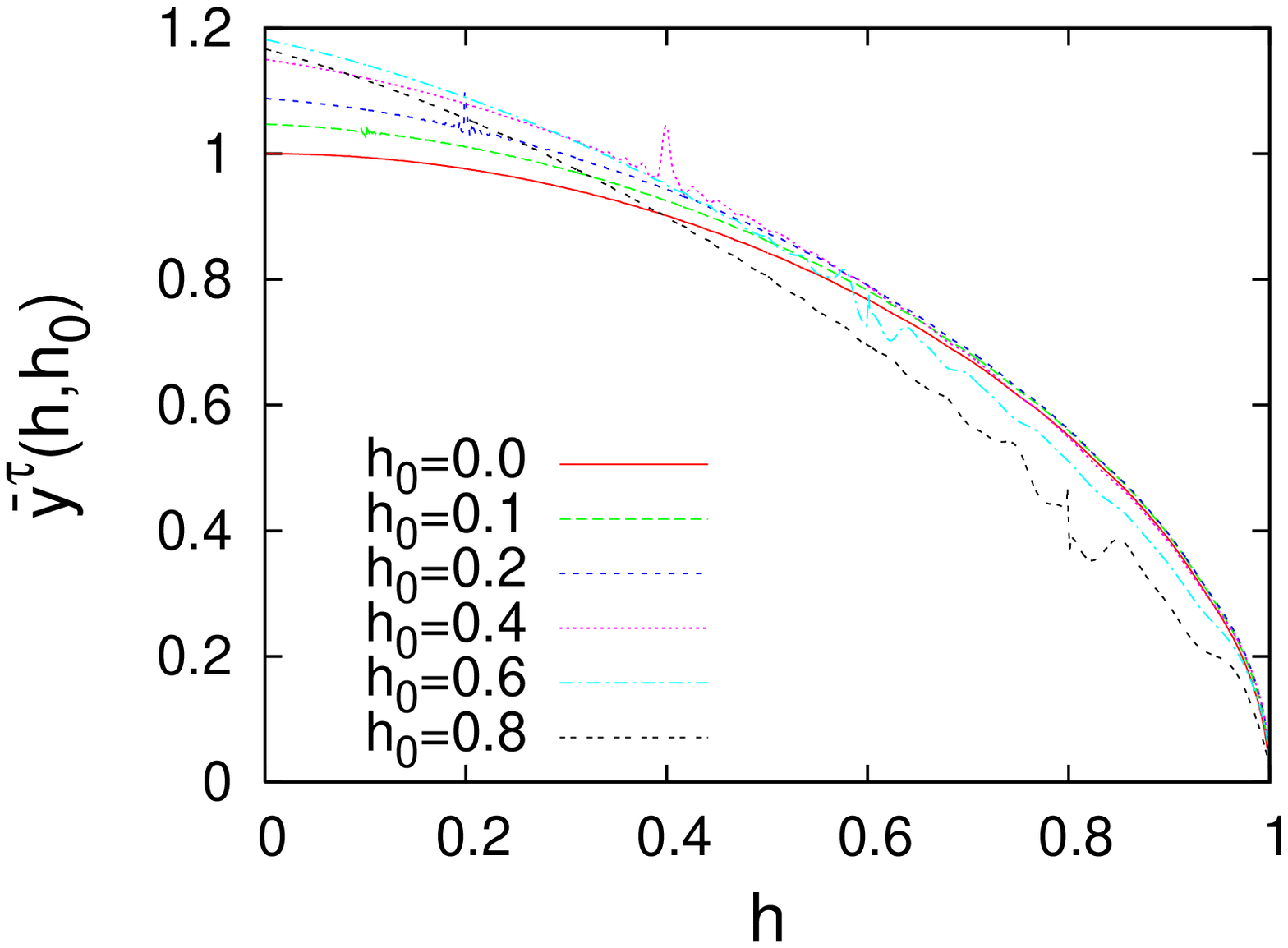}
\end{center}
\vskip -.5cm
\caption{
\label{fig_3} (Color online)
{\bf Left:} The ratios $y^\xi=\Delta\tilde\xi(h)/\Delta\tilde\xi(0)$ 
and $y^\tau=\Delta\tilde\tau(h)/\Delta\tilde\tau(0)$ 
for a quench from $h_0=0$ as a function of $h$. The curve 
$y^{\xi'}$ derives from the exactly known form for $\xi(h)$, 
see text. {\bf Right:} The
ratios 
$\overline{y}^\tau(h,h_0)=
\Delta\tilde\tau(h,h_0)/\Delta\tilde\tau(0,0) 
=\tau(h,h_0)\cdot h (h-h_0)^2/\pi - (1-h_0)^2/2$ 
for various $h_0$ as a function of $h$.}
\end{figure}

To obtain information about $\tau(h,h_0)$ away from the singularities
we consider a quench from the fully ordered initial state ($h_0=0$) 
first. A quench into the disordered phase ($h\ge1$)
yield to high numerical accuracy 
$\tau(h \ge 1,h_0=0)=\pi/2$, i.e.\ independent of $h$.
For a quench into the ordered phase ($h\le1$)
we introduce $\tilde{\tau}(h,h_0=0)=h^3\tau(h,h_0=0)$
to get rid of the singularity at $h=0$.
In the limit $h\to0$ we obtain $\tilde{\tau}(h=0,h_0=0)=3\pi/2$,
and for $h>0$ we consider the ratio:
$y^{\tau}(h)=\Delta\tilde\tau(h)/\Delta\tilde\tau(0)$
with $\Delta\tilde\tau(h)=\tilde\tau(h)-\tilde\tau(1)$
and compare it with a similar expression for
the correlation length 
$y^{\xi}(h)=\Delta\tilde\xi(h)/\Delta\tilde\xi(0)$
with $\Delta\tilde\xi(h)=\tilde\xi(h)-\tilde\xi(1)$,
where $\tilde{\xi}(h)=\xi(h)h^2$.
The two ratios $y^\tau(h)$ and $y^\xi(h)$, as shown in
Fig.\ref{fig_3}a, are almost indistinguishable. Since 
$\xi(h)=-1/\log((1+\sqrt{1-h^2})/2)$ is known
exactly\cite{sengupta}, the relaxation time for a quench from an
ordered initial state ($h_0=0$) can therefore be estimated
very accurately, if not exactly, by the relation 
$y^{\tau}(h)=y^{\xi}(h)$.

Starting from a partially ordered initial state ($0<h_0<1$) we define $\tilde{\tau}(h,h_0)=h(h-h_0)^2\tau(h,h_0)$
and find to
high numerical accuracy that the limiting value
at $h=1$ is given by: $\tilde{\tau}(h=1,h_0)=\pi(1-h_0)/2$. Away from
$h=1$ we study the ratio
$\overline{y}^{\tau}(h,h_0)=\Delta\tilde\tau(h,h_0)/\Delta\tilde\tau(0,0)$ with
$\Delta\tilde\tau(h,h_0)=\tilde\tau(h,h_0)-\tilde\tau(1,h_0)$ which is
identical to $y^{\tau}(h)$ for $h_0=0$ and which is plotted in
Fig.\ref{fig_3}b for different values of $h_0$. The curves for all
values of $h_0$ are quite close to each other, and at $h=1$
they all have a singularity, $\sim \sqrt{1-h}$. Therefore
one obtains a very good estimate for the relaxation time
from $\tilde{\tau}(h,h_0)$ by $\overline{y}^\tau(h,h_0)\approx
y^\tau(h)=y^{\xi'}(h)$, which is given in an analytical form (see above).

The thermal-like stationary state can be characterized by an effective
temperature $T_{eff}$ \cite{rossini} which is defined through the
condition, that the relaxation time in the stationary state after a
quench, $\tau(h,h_0)$, and the equilibrium correlation time at
temperature $T=T_{eff}$, $\tau_T(h,T)$, are identical. Using the
analytic result at the critical point\cite{deift}:
$\tau_T(h=1,T)=8/(\pi T)$ we arrive at
$T_{eff}(h_0,h=1)=16(1-h_0)/\pi^2$, which is compatible with the
numerical data in Ref.\cite{rossini}. In the ferromagnetic phase,
$h<1$, and in the limit $T \ll \Delta(h)$, $\Delta(h)$ being the gap,
the relaxation time is given by \cite{sachdev_young}: $\tau_T(h<1,T)
\approx (2/(\pi T)) e^{\Delta/T}$, which for
$|h-h_0| \ll 1$ leads to: $T_{eff} \approx -\Delta(h)/(2
\ln|h-h_0|)$.

To summarize we have identified different regimes in the
non-equilibrium relaxation of the magnetization profiles of the
quantum Ising chain with boundaries, which can be explained in terms
of quasi-particles that are reflected at the surfaces.  For sites at
or near the surface non-thermal behavior is observed, manifested by a
power-low relaxation form. For bulk sites a cross-over to thermal
behavior is found, with exponentially decaying correlations, defining
a relaxation time and a correlation length that is
identical in semi-infinite and in infinite systems and which obey
presumably exact relations conjectured on the basis of the numerical
data. In a finite system an exponentially fast reconstruction of the
local magnetization is observed, involving a time-scale, $\tau'$,
and characterizing an approximately periodic dynamics.

Several results for observables displaying thermal behavior
in the bulk are expected to be valid also in other, even
non-integrable spin chains: Absence of thermalization at the
boundaries, identity of correlation time and length in infinite and
semi-infinite systems and an exponentially fast
reconstruction in finite systems.

\begin{acknowledgments}
This work has been supported by the Hungarian National Research Fund
under grant No OTKA K62588, K75324 and K77629 and by a
German-Hungarian exchange program (DFG-MTA).
\end{acknowledgments}

\end{document}